\documentclass[journal]{IEEEtran}
\usepackage{amsmath,amsfonts,amssymb,mathrsfs}
\usepackage{algorithmic}
\usepackage{algorithm}
\usepackage{array}
\usepackage{booktabs}
\usepackage{subcaption}
\usepackage{textcomp}
\usepackage{hyperref}
\usepackage{stfloats}
\usepackage{url}
\usepackage{xcolor}
\usepackage{verbatim}
\usepackage{graphicx}

\begin{document}

\title{A New Paradigm of User-Centric Wireless Communication Driven by Large Language Models}
\author{Kuiyuan Ding, Caili Guo, \textit{Senior Member, IEEE}, Yang Yang, \textit{Senior Member, IEEE}, Wuxia Hu and Yonina C. Eldar, \textit{Fellow, IEEE}}

\maketitle
\footnotetext{K. Ding, Y. Yang and W. Hu are with the Beijing Key Laboratory of Network System Architecture and Convergence, School of Information and Communication Engineering, Beijing University of Posts and Telecommunications, Beijing 100876, China (e-mail: dingkuiyuan@bupt.edu.cn; yangyang01@bupt.edu.cn; wuxiahu@bupt.edu.cn).

C. Guo is with the Beijing Laboratory of Advanced Information Net works, School of Information and Communication Engineering, Beijing University of Posts and Telecommunications, Beijing 100876, China (e-mail: guocaili@bupt.edu.cn).

Yonina C. Eldar is with the Faculty of Mathematics and Computer Science, Weizmann Institute of Science, Rehovot 7610001, Israel. (e-mail:yonina.eldar@weizmann.ac.il)
}
\begin{abstract}
The next generation of wireless communications seeks to deeply integrate artificial intelligence (AI) with \textcolor{black}{user-centric} communication networks, with the goal of developing AI-native networks that more accurately address user requirements. The rapid development of large language models (LLMs) offers significant potential in realizing these goals. However, existing efforts that leverage LLMs for wireless communication often overlook the considerable gap between human natural language and the intricacies of real-world communication systems, thus failing to fully exploit the capabilities of LLMs. To address this gap, we propose a novel LLM-driven paradigm for wireless communication that innovatively incorporates the nature language to structured query language (NL2SQL) tool. Specifically, in this paradigm, user personal requirements is the primary focus. Upon receiving a user request, LLMs first analyze the user intent in terms of relevant communication metrics and system parameters. Subsequently, a structured query language (SQL) statement is generated to retrieve the specific parameter values from a high-performance real-time database. 
We further utilize LLMs to formulate and solve an optimization problem based on the user request and the retrieved parameters. The solution to this optimization problem then drives adjustments in the communication system to fulfill the user’s requirements. To validate the feasibility of the proposed paradigm, we present a prototype system.  
In this prototype, we consider user-request centric semantic communication (URC-SC) system in which a dynamic semantic representation network at the physical layer adapts its encoding depth to meet user requirements, such as improved data transmission quality or reduced data transmission latency.
Additionally, two LLMs are employed to analyze user requests and generate SQL statements, respectively. Simulation results demonstrate the effectiveness of the prototype in personalizing and addressing user requirements through the \textcolor{black}{user-centric wireless communication} paradigm. Code: \textcolor{magenta}{\url{https://github.com/echojayne/URC-SC.git}}.
\end{abstract}


\section{Introduction}\label{section:intro}

\begin{figure*}[t!]
    \centering
    \includegraphics[width=\linewidth]{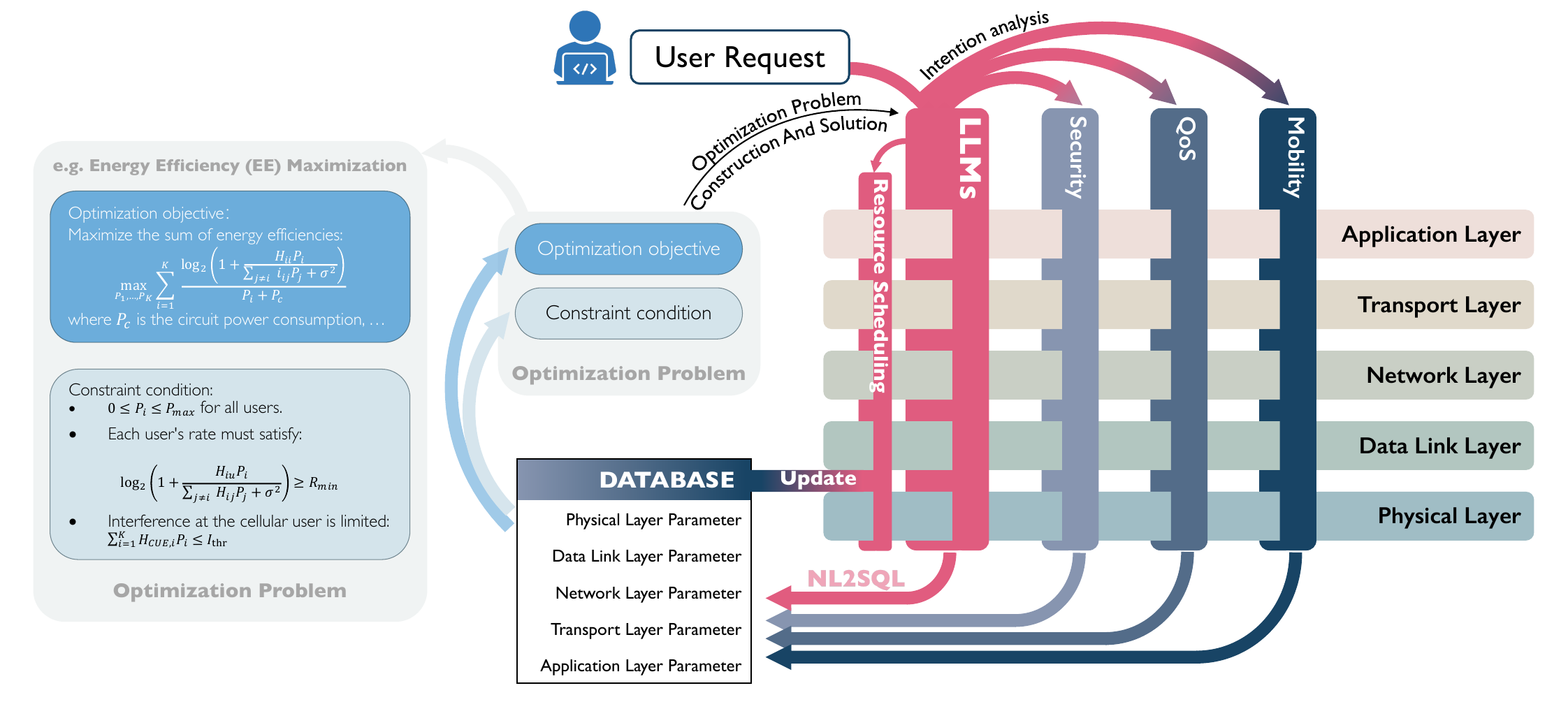}
    \caption{The proposed \textcolor{black}{user-centric wireless communication} paradigm.}
    \label{fig:framework}
\end{figure*}
The future generation of wireless communications, specifically the sixth generation (6G), aims to deeply integrate artificial intelligence (AI) with communication networks to achieve AI-native network design. This integration can enhance interactions between humans and machines and meet user requirements more precisely. Unlike leveraging AI as external modules, AI-native network design means AI will be embedded at every layer of the communication architecture including using AI to manage resources, optimize network slices, and predicatively allocate bandwidth to meet user requirements in real-time.

In the past decades, AI technologies have developed rapidly, playing an increasingly important role in wireless communication systems. Traditional AI approaches, such as machine learning (ML) and deep learning (DL), have been successfully applied to various tasks within communication networks, including channel estimation
\cite{AI-CSI-1, AI-CSI-2}, interference management
\cite{AI-Interference-2, AI-Interference-3} 
and resource allocation
\cite{AI-resource-1, AI-resource-3}. These approaches typically rely on carefully designed feature extraction and model training pipelines tailored to specific communication scenarios. For instance, supervised learning techniques have been employed to optimize physical-layer parameters, while reinforcement learning (RL) has been utilized for dynamic spectrum allocation and network slicing.
However, the effectiveness of traditional AI technologies in communication systems is often constrained by their reliance on domain-specific datasets and task-specific algorithms. These methods require substantial expert intervention to design models and features, limiting their scalability and adaptability to diverse communication scenarios. Furthermore, the fragmented nature of traditional AI systems, where different models are developed for individual tasks, poses challenges for seamless integration and cross-layer optimization within the network.

Starting with ChatGPT-3 , developed by OpenAI, large language models (LLMs) officially entered the public spotlight, marking a paradigm shift in AI applications across various domains. These models, especially the most advanced models, such as DeepSeek-R1, OpenAI o3-mini and Grok3 etc., characterized by their massive scale with billions of parameters, exhibit exceptional performance in natural language processing (NLP) tasks such as question answering, natural language understanding, and text generation. By leveraging extensive pre-training on massive natural language corpora, LLMs demonstrate remarkable contextual understanding and creativity, far surpassing the capabilities of earlier AI models. Their ability to comprehend and generate human-like text allows them to effectively capture human intent and respond accordingly, introducing unprecedented opportunities for enhancing communication networks.

Unlike traditional AI approaches that rely on domain-specific datasets and task-specific algorithms, LLMs are general-purpose and excel in understanding and processing unstructured data. This makes them particularly suitable for interpreting user intents expressed in natural language, satisfying a more intuitive and personalized user requirements. For instance, LLMs can dynamically translate user needs into actionable network-level configurations, optimizing key parameters such as latency, throughput, and reliability. This transition from task-specific AI models to more unified and adaptable AI-driven architectures represents a significant evolution in communication system design.
By embedding LLMs as foundational models in communication systems, the integration of native AI becomes more feasible, addressing diverse user needs while enhancing interaction quality. These capabilities align seamlessly with the vision of 6G wireless communication systems, where AI-native network designs are expected to manage resources, optimize network slices, and predicatively allocate bandwidth based on real-time demands. LLMs’ ability to analyze unstructured human input and adapt network behavior accordingly makes them a powerful tool for realizing intelligent, user-centric communication systems.

Some existing studies have conducted preliminary explorations into the application forms of LLMs in wireless communication systems. 
In \cite{lee2024llmempoweredresourceallocationwireless}, the authors explored the integration of LLMs into wireless communication systems by proposing an LLM-based resource allocation scheme.
In \cite{LLM_Twin}, the authors presented a LLM-empowered framework for digital twins networks called LLM-Twin, featuring a mini-giant model collaboration scheme for efficient deployment and a semantic-level, high-efficiency, and secure communication models. 
In \cite{guo2024semanticimportanceawarecommunicationssemantic}, the authors proposed an understanding-level semantic communications scheme, which transforms visual data into human-intelligible natural language descriptions using an image caption neural network and refines them with a LLM for importance quantification and semantic error correction. 
These initial efforts have inspired innovative methodologies for integrating LLMs into communication systems, primarily addressing communication issues by utilizing LLMs to interpret natural language descriptions of these problems and subsequently generate appropriate solutions. 

 Nevertheless, a significant gap persists regarding the practical integration of LLMs into real-world communication systems, particularly due to differences in data formats processed by LLMs and those required by communication systems. Specifically, LLMs are inherently text-driven and optimized for processing natural language, whereas communication systems predominantly operate with structured data formats, numerical values, and complex signal representations, including modulation schemes, channel states, and network parameters. Consequently, there remains a pronounced discrepancy between the natural language inputs managed by LLMs and the intricate, parameterized configurations essential in practical communication environments.
\begin{figure*}[ht!]
    \centering
    \includegraphics[width=\linewidth]{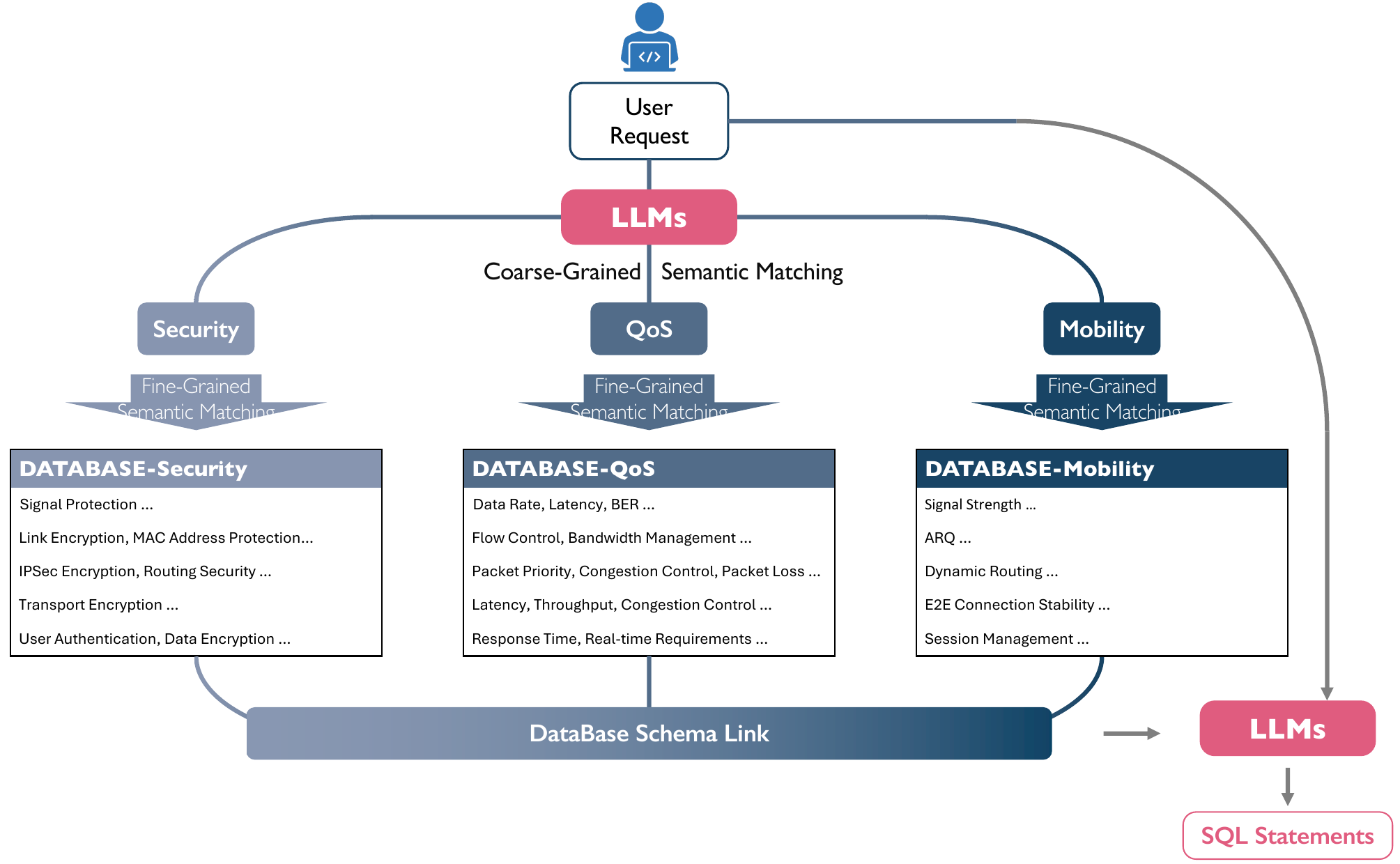}
    \caption{The framework of NL2SQL tool.}
    \label{fig:nl2sql}
\end{figure*}

In this article, we introduce a new paradigm for \textcolor{black}{user-centric} wireless communications enabled by LLMs, which focuses on user requests and leverages the power of LLMs to bridge the gap between human natural language requests and the intricate parameterized configuration of real world communication systems, utilizing the natural language to structured query language (NL2SQL) tool \textcolor{black}{which is used to simplify database interactions by allowing users to retrieve and manipulate data using everyday language instead of complex structured query language (SQL) queries}. This novel paradigm represents a significant shift in the way communication resources are scheduled and optimized, placing user-request-centric intelligence at the core of the system. Specifically, we propose a paradigm where LLMs serve as an intermediary layer between the user natural language input and the technical requirements of the communication system. The user, serving as the central entity, may issue requests in natural language, \textcolor{black}{such as seeking lower transmission latency or higher transmission quality}. Subsequently, LLMs interpret these requests and translate them into actionable communication metrics. These translations take the form of SQL statements, which are executed on a database containing key parameters of the communication system’s multi-layer architecture. Following this, LLMs return optimization objectives and constraints based on both the user’s needs and the current network conditions, effectively formulating the optimization problem. Through the inherent intelligence of LLMs, this optimization problem is solved, providing a dynamic scheduling solution that aligns communication resources with the user’s request while continuously updating the system’s data storage to reflect these changes. To demonstrate the feasibility and effectiveness of this paradigm, we have developed a prototype system that implements this paradigm. In this prototype, we consider a user-request centric semantic communication (URC-SC) system.
The system enables intuitive, user-request centric control of network transmission behavior, responding more effectively to users' personalized communication needs.
Experimental results validate the effectiveness of this new paradigm. When a user requests enhanced data transmission quality five consecutive times, the transmission quality—represented by the classification accuracy of the URC-SC system—improves from 68.99\% to 93.80\% at a SNR of 20 dB under AWGN channel conditions. Concurrently, this enhancement leads to an increase in data transmission latency from 105.3757 ms to 167.1618 ms per image.

When the user requests higher data transmission quality twice, the accuracy increases from 76.46\% to 85.91\% at an SNR of 20 dB, and when the user requests lower data transmission latency twice, the latency decreases from 117.7153 ms to 93.2484 ms per image.



\section{User-Centric Wireless Communication Enabled by LLMs}\label{section:framework}

In this section, we first introduce the framework of the proposed \textcolor{black}{user-centric wireless communication} paradigm, followed by a discussion of the key design aspects of the framework.
\subsection{Basic Framework}
Fig. \ref{fig:framework} illustrates the framework of proposed \textcolor{black}{user-centric wireless communication} paradigm. The entire system uses LLMs as an intermediary to bridge the gap between the user and the communication system. When a user submits a request, the LLMs analyze the user’s intent and translates the request into specific demands related to the current communication system’s metrics, such as \textit{Mobility}, \textit{Quality of Service (QoS)}, and \textit{Security}. These metric requests are further analyzed into specific parameters for the various layers of the communication system. The relevant parameters are categorized into optimization objectives and constraint conditions based on the user’s request and the current network limitations.

To manage these parameters, we maintain a database that stores the necessary values. LLMs generate SQL queries to retrieve the specific parameter values based on the user’s requirements. The formulated optimization problem is then solved by LLMs, utilizing the queried values. The solution to the optimization problem is used to schedule communication resources in a manner that meets the user’s needs. Finally, the modified parameters are updated in the database to reflect the changes.
\subsection{Key Designs}
\begin{figure*}
    \centering
    \includegraphics[width=\linewidth]{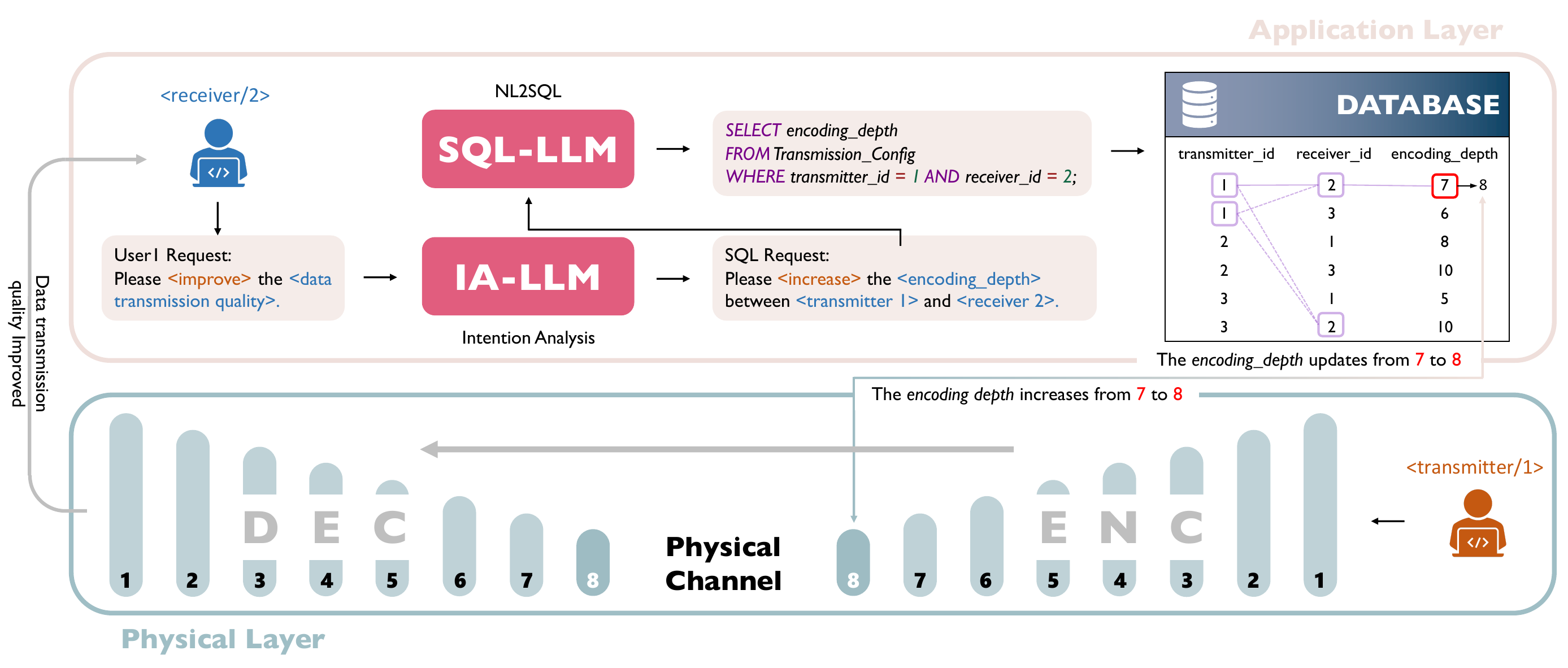}
    \caption{A prototype system of \textcolor{black}{user-centric wireless communication} paradigm.}
    \label{fig:system}
\end{figure*}

\subsubsection{Nature Language to SQL}
The primary challenge in employing LLMs for wireless communications lies in enabling them to understand user requests expressed in natural language while simultaneously sensing and capturing the real-time, dynamic status of the communication system.
In the proposed paradigm, we introduce an innovative approach in wireless communication that integrates an NL2SQL tool to bridge the gap between LLMs and wireless communication systems as shown in Fig. \ref{fig:nl2sql}. Our architecture maintains high-frequency, real-time databases that store system parameters reflecting the current state of the network. 
When a user submits a request, a coarse-grained semantic matching process first categorizes it into broad system metrics such as Security, QoS, and Mobility. This initial categorization is then refined through fine-grained semantic matching within each metric category, thereby identifying specific, relevant system parameters across different layers of the communication system.
To facilitate real-time interaction and data retrieval based on the analyzed intent, the system leverages NL2SQL. The continuously updated real-time databases provide the current status of system parameters. Once the LLMs analyze the user’s request and pinpoint the necessary parameters, they extract the relevant database schema linkage information to generate SQL queries via NL2SQL. 
These queries are executed against the real-time databases, enabling the LLMs to retrieve precise, up-to-date parameter values corresponding to the user’s original request.
This framework allows LLMs to focus on understanding and mapping user needs to system controls, while the database efficiently manages real-time data updates. For instance, when a user requests specific QoS guarantees, the LLM translates that request into SQL queries to fetch current metrics such as bandwidth utilization, latency, and other performance indicators. Similarly, for security-related queries, the LLM retrieves information on encryption settings or user authentication policies to ensure that the network satisfies the required security standards.

Overall, this framework not only offers a precise and flexible means of meeting diverse user requirements but also reduces the computational burden on LLMs. By offloading real-time data processing to the database layer, the system can rapidly adapt to changing conditions and deliver targeted control and scheduling in wireless networks, thereby paving the way for more intelligent and responsive network management.

\subsubsection{Optimization Problem Construction and Solution}
The user request ultimately reflects the need to adjust system parameters across various layers. In general, it is not feasible to allocate all resources to a single user to fully meet its demands. Instead, there must be a balance between user requests and available system resources. Therefore, we can formulate an optimization problem aimed at satisfying the maximum possible user needs within the current resource constraints. Specifically, we categorize the retrieved parameters into optimization objectives and constraint conditions based on the user’s request and the limitations of the network. LLMs are then utilized to construct and solve the optimization problem. Finally, the system adjusts its parameters according to the solution.

The communication system is scheduled based on the solution to the optimization problem. Hence, the ability to correctly solve the optimization problem depends on whether the user request can be accurately conveyed to the system. Recent advancements in the mathematical capabilities of LLMs, particularly with the development of chain-of-thought (CoT) reasoning, have greatly enhanced their problem-solving abilities. Notable models such as DeepSeek-R1, developed by deepseek team, ChatGPT o1 Pro, developed by OpenAI, and Claude 3.5 Sonnet, developed by Anthropic, showcase these advancements. However, these models still face limitations in solving complex optimization problems with perfect accuracy. Nevertheless, in practical communication systems, the number of parameters that can be adjusted is relatively small, and the optimization problems we construct  are typically simple enough for LLMs to solve correctly. This enables the proper scheduling of system resources to meet user needs effectively.
\section{Prototype System}\label{section:implementation}

\begin{figure*}[ht!]
    \centering
    \begin{subfigure}[b]{0.48\textwidth}
        \centering
        \includegraphics[width=\textwidth]{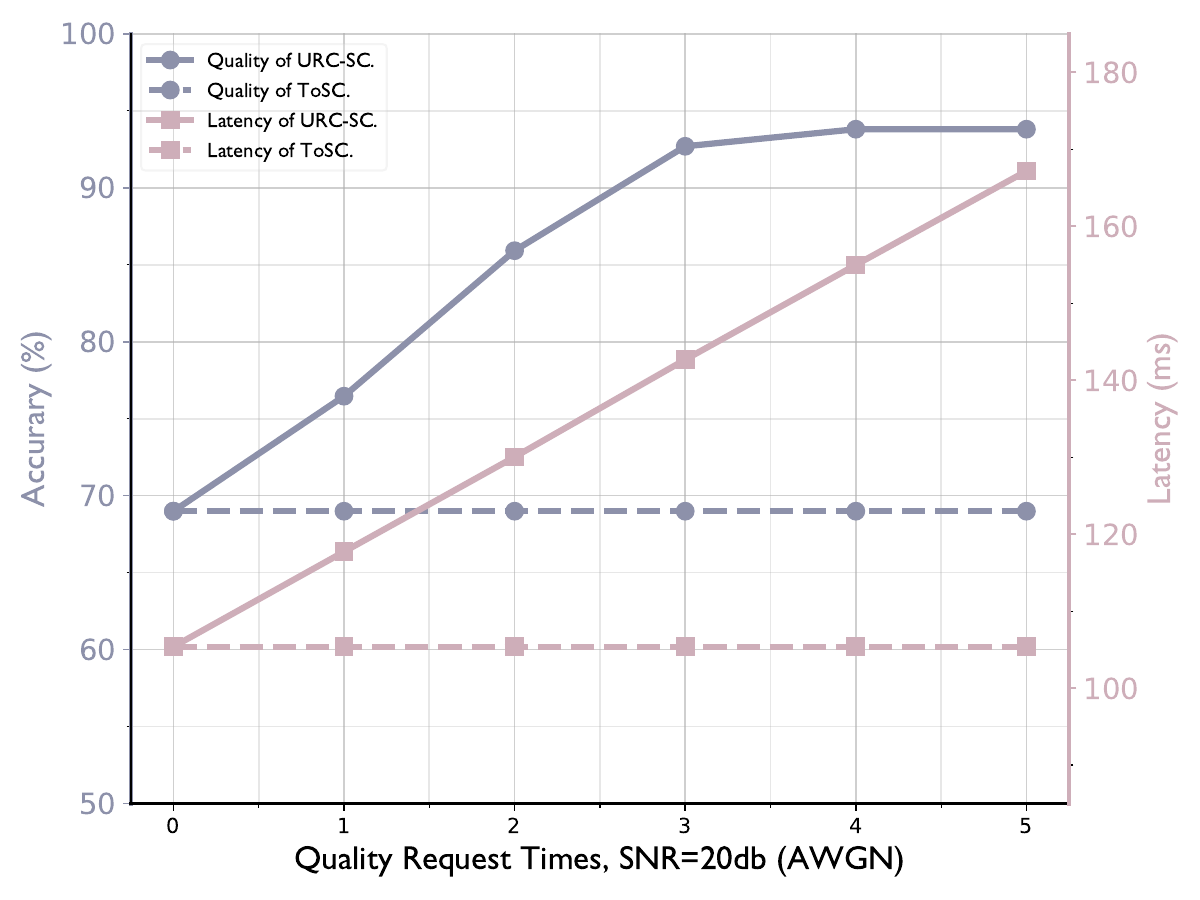}
        \caption{Request for Data Transmission Quality}
        \label{fig:AWGN}
    \end{subfigure}
    \hfill
    \begin{subfigure}[b]{0.48\textwidth}
        \centering
        \includegraphics[width=\textwidth]{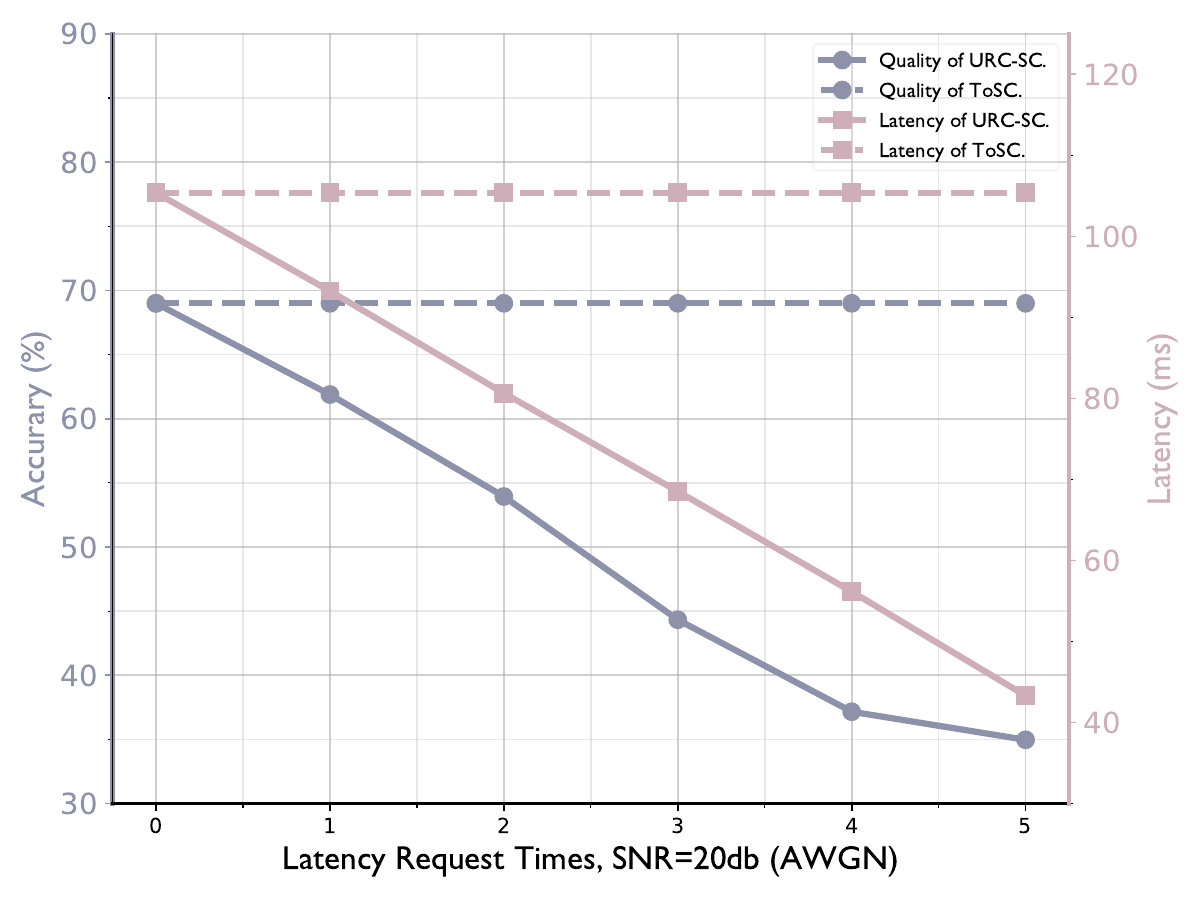}
        \caption{Request for Data Transmission Latency}
        \label{fig:Rayleigh}
    \end{subfigure}
    \caption{The response of the communication system to user requests.}
    \label{fig:results}
\end{figure*}
In this section, we first present a prototype system as a simple implementation of the proposed \textcolor{black}{user-centric wireless communication} paradigm in Section \ref{section:framework}, followed by a discussion of the results obtained from the prototype.
\subsection{Implementation}
Fig. \ref{fig:system} illustrates the prototype system of the proposed \textcolor{black}{user-centric wireless communication} paradigm. In this prototype, we simplify the system by focusing solely on the application and physical layers. Specifically, we consider a 
URC-SC system at the physical layer. In the application layer, the user submits requests related to system performance. The semantic codec at the physical layer is represented as a dynamic semantic representation network with a variable encoding depth. The deeper the encoding depth, the better the task performance of the system, although it requires more time to extract the semantic features. For simplicity, we focus on user requests related to data transmission such as task performance or transmission latency.

For example, when the user submits the request: “Please improve the data transmission quality,” the LLM used for intention analysis (IA-LLM) first analyzes which parameter—specifically the encoding depth in this case—is related to the request and how to modify it (i.e., increase or decrease the encoding depth). The IA-LLM then generates a SQL query: “Please increase the encoding depth between transmitter 1 and receiver 2.” This query is processed by an NL2SQL LLM (SQL-LLM) to generate the corresponding SQL statement. The SQL statement is executed to retrieve the current encoding depth. If the current encoding depth has not yet reached the system’s maximum, it is increased by one, in accordance with the user’s intent. Finally, the encoding depth in the database is updated, and the user’s requirements is satisfied.

\subsubsection{Dynamic Semantic Representation Network}
The dynamic semantic representation network is essentially a Vision Transformer (ViT) with a variable encoding depth, which can dynamically select the output layer based on a parameter. Deeper encoding layers produce more compact semantic features with reduced redundancy, but require more computation time. In contrast, shallower encoding layers generate sparser semantic features with higher redundancy, but are faster to compute.

\subsubsection{Optimization Problem}
Given the simplicity of the system, there is no need to formulate a complex optimization problem. The only variable in this context is the encoding depth. When the user requests an improvement in data transmission quality, the encoding depth is increased by one layer. Conversely, when the user requests a reduction in data transmission latency, the encoding depth is decreased by one layer. This process continues until the system reaches its maximum or minimum encoding depth.

\subsection{Simulation Settings}
\subsubsection{Task and Dataset}
We consider a classification task at the physical layer, where the transmitter extracts and transmits the semantic features of an image, and the receiver decodes these features to generate the target of the task, namely the image’s category. We train and evaluate the performance of the system using the CIFAR-10 dataset, which consists of 60,000 32x32 color images, distributed across 10 classes with 6,000 images per class.
\subsubsection{Foundation Moldes}
We utilize Qwen2.5, developed by the Qwen Team, as the foundation models for our system. Specifically, we employ Qwen2.5-7B-Instruct as the IA-LLM and Qwen2.5-Coder-7B-Instruct as the SQL-LLM. These models are well-suited for performing intention analysis and NL2SQL tasks with high precision.
\subsubsection{Baselines}
We adopt a conventional task-oriented semantic communication (ToSC) system as the baseline, which features a structure similar to that described in \cite{deepJSCC} incorporates a deep joint source and channel coding (deep-JSCC) encoder at the transmitter and a corresponding deep-JSCC decoder at the receiver.

\subsection{Simulation Results}
Fig. \ref{fig:results} presents the simulation results illustrating how the proposed URC-SC system responds to user requests. As shown in Fig. \ref{fig:AWGN}, when a user requests enhanced data transmission quality five consecutive times, the transmission quality—represented by the classification accuracy of the URC-SC system—improves significantly from 68.99\% to 93.80\% at a signal-to-noise ratio (SNR) of 20 dB under additive white Gaussian noise (AWGN) channel conditions. Concurrently, this enhancement leads to an increase in data transmission latency from 105.3757 ms to 167.1618 ms per image. In contrast, the conventional ToSC system lacks the capability to receive and respond to user requests, thus failing to provide any improvement in response to user demands. Similar trends are observed when users request lower data transmission latency. As illustrated in Fig. \ref{fig:Rayleigh}, when users request reduced latency five consecutive times, the URC-SC system successfully decreases the transmission latency from 105.3757 ms to 43.3145 ms per image at an SNR of 20 dB under AWGN channel conditions. However, this improvement in latency comes with a trade-off, as the data transmission quality concurrently declines from 68.99\% to 34.97\% per image. The conventional ToSC system still fails to provide any improvement in response to user demands.

These results demonstrate that the prototype effectively understands user intent and satisfies user needs, a capability the conventional ToSC system lacks. Powered by LLMs, network transmission behavior can be dynamically adjusted according to user requests, thereby significantly enhancing the overall user experience. The simulation results preliminarily validate the feasibility of personalizing wireless communication systems and effectively meeting diverse user requirements through the \textcolor{black}{user-centric wireless communication} paradigm.

\subsection{More Comparison}
\begin{figure}[tb]
    \centering
    \includegraphics[width=\linewidth]{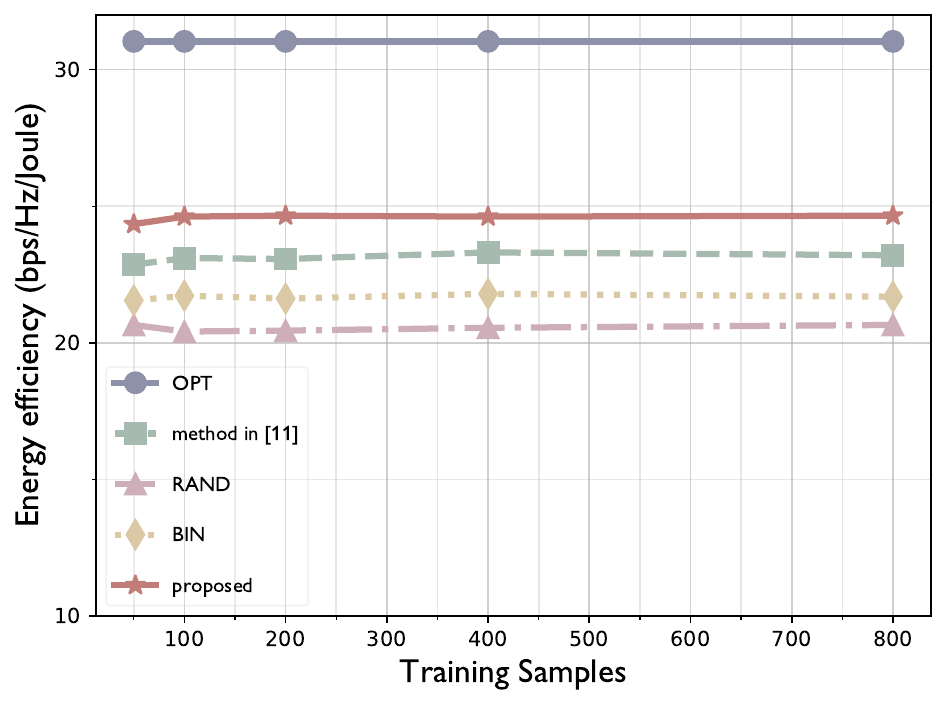}
    \caption{Energy efficiency vs. number of training samples.}
    \label{fig:LLM_RA}
\end{figure}
\textcolor{black}{We further compare the proposed user-centric wireless communication framework with the LLM-based resource allocation scheme presented by Lee et al.~\cite{lee2024llmempoweredresourceallocationwireless}, where the authors explored a resource allocation schema empowered by LLMs in wireless communication systems. In our proposed user-centric paradigm, key wireless communication parameters, such as bandwidth and channel gain, are explicitly stored in a structured database. The resource allocation task, specifically the power allocation problem, is formulated as a mathematical optimization problem, which is directly provided to the LLM to generate the solution.}

\textcolor{black}{As illustrated in Fig.\ref{fig:LLM_RA}, under identical configuration settings to those used in \cite{lee2024llmempoweredresourceallocationwireless}, our proposed user-centric wireless communication framework achieves an energy efficiency improvement exceeding 5\% compared to the method of Lee et al. \cite{lee2024llmempoweredresourceallocationwireless}. This enhancement arises primarily due to the clearer formulation of the power allocation problem as a mathematically structured optimization task. Unlike the in-context learning method adopted in \cite{lee2024llmempoweredresourceallocationwireless}, our approach facilitates LLM comprehension by explicitly defining wireless communication system parameters and their physical significance, thereby enabling more effective optimization.}

\section{Open Issues and Future Research Directions}\label{section:issues}

\subsection{Open Issues}
\subsubsection{Inherent Issues of LLMs}
Since the paradigm we propose is LLM-driven, it is inherently influenced by the limitations and challenges associated with LLMs. First, LLMs require significant computational resources, which not only demand substantial processing power from the communication system but also contribute to high power consumption. Second, the inference delay of large models is directly tied to their computational demands. The extensive computation required by LLMs leads to high inference latency, which in turn reduces the system’s responsiveness to user requests, ultimately affecting the user experience. Third, LLMs are inherently prone to “hallucination,” meaning that they may generate unreliable outputs. This issue could result in the system receiving incorrect or fabricated requests, undermining the system’s reliability. Finally, data privacy concerns arise due to the ability of LLMs to access and process all state information of the communication system, potentially leading to data privacy breaches. These challenges are intrinsic to LLMs and must be addressed in order to build a responsive, accurate, and secure system.
\subsubsection{Multi-user Conflict and Concurrency}
In a real communication system, where services are provided to multiple users, requests from different users may conflict with each other. Additionally, when multiple users request the same resource, concurrency issues may arise. Furthermore, frequent requests from multiple users could lead to frequent changes in the system, potentially destabilizing it and negatively affecting the user experience. Therefor, resource scheduling and allocation under multi-user requests remains a challenging problem. 

\subsubsection{NL2SQL}
Fig. \ref{fig:nl2sql} illustrates the proposed NL2SQL framework, which is designed to bridge the gap between LLMs and wireless communication systems. However, the NL2SQL task is inherently complex, and these complexities are further amplified when queries span multiple tables in a database. In such cases, the necessity for nested structures or subqueries increases the risk of generating incorrect or inefficient SQL statements. Even minor inaccuracies in the translation process can lead to significant data retrieval errors, thereby potentially compromising the reliability of the wireless communication system. Moreover, the computational overhead associated with processing these intricate queries may result in increased inference latency, ultimately undermining the system’s ability to deliver real-time responses. Addressing these open issues is essential for advancing NL2SQL technology and ensuring its capability to meet the dynamic and complex demands of modern wireless communication networks.
\subsection{Future Research Directions}
\subsubsection{Native AI Design}
There are several challenges, as discussed in the Open Issues section, that may arise when directly applying the proposed \textcolor{black}{user-centric wireless communication} paradigm to current communication systems. To maximize the effectiveness of LLMs, a promising approach is to design a native AI-powered communication system. The prototype system presented in Section \ref{section:implementation} serves as an example of such an approach. Native AI design allows LLMs to be deeply integrated into all aspects of the system, facilitating more seamless and efficient management of communication processes. Furthermore, the optimal native AI design for wireless communication involves embedding multiple AI modules throughout the system to perform various tasks, such as source coding, channel prediction, and more. These AI modules would be managed by LLMs to meet the personalized demands of individual users better.

Future research in this direction could focus on the development of AI-powered communication protocols tailored to specific use cases, such as real-time data transmission, low-latency communication, or resource-constrained environments. Additionally, integrating multi-agent AI systems into communication networks could further enhance the scalability and flexibility of the system, enabling dynamic adaptation to varying user demands and network conditions. To achieve this, several key challenges must be addressed in future research:

\begin{itemize}
\item Which components of communication systems can be replaced by AI modules to achieve optimal performance?
\item How can LLMs efficiently manage these AI modules in the native AI designs of communication systems?
\item How can the trade-off between power dissipation and system performance be balanced in native AI designs?

\end{itemize}
\subsubsection{Human-in-the-loop for \textcolor{black}{User-Centric Wireless Communication}}
In Section \ref{section:framework}, we extensively discussed user intention analysis and highlighted the challenges associated with it in the Open Issues section. The primary gap, however, lies in the translation of user requests into specific parameters of the communication system. The key challenge, therefore, is enabling LLMs to effectively understand user needs in the context of communications. One promising approach to address this challenge is the Human-in-the-loop (HITL) methodology \cite{wu2022survey}. HITL is a collaborative approach that integrates human expertise and input throughout the lifecycle of machine learning (ML) and AI systems. Its primary aim is to enhance the accuracy, reliability, and adaptability of ML systems by leveraging the unique capabilities of both humans and machines \cite{google_hitl}.

In the context of \textcolor{black}{user-centric wireless communication} systems, HITL can facilitate the long-term learning of the relationship between user requests and system parameters, enabling LLMs to better understand and fulfill user demands. Furthermore, HITL can help LLMs acquire deeper expertise in wireless communications, which in turn would enhance their ability to manage and optimize system performance more efficiently. Given this potential, it is crucial to explore how HITL technologies—such as SFT, Reinforcement Learning from Human Feedback (RLHF), Direct Preference Optimization (DPO), Proximal Policy Optimization (PPO) and Group Relative Policy Optimization (GRPO)\cite{GRPO}, which has been proven to have a significant impact through the DeepSeek-R1 model—can empower \textcolor{black}{user-centric wireless communication} systems. These techniques hold promise in refining LLMs’ ability to adapt to dynamic user needs and optimize system-level parameters, contributing to the overall efficiency and user experience in communication networks.
\section{Conclusion}

In this article, we have proposed a novel paradigm for wireless communication driven by LLMs. Specifically, we have introduced a \textcolor{black}{user-centric wireless communication} paradigm in which the user is considered the central entity, and LLMs have been employed to schedule communication resources utilizing NL2SQL tools based on user requests. We have then presented a prototype system of this paradigm that implements a user-request centric semantic communications system. In the proposed prototype, an IA-LLM and a SQL-LLM have been used to analyze user intent and generate SQL statements to retrieve current system parameters, respectively. A dynamic semantic representation network at the physical layer has adapted its encoding depth based on user demands. Subsequently, we have evaluated the system’s efficiency in satisfying user needs through experimental validation. Finally, we have discussed the open issues associated with \textcolor{black}{user-centric wireless communication}s and outlined potential future research directions in this field.


\bibliographystyle{IEEEtran}
\bibliography{Reference}

\end{document}